\documentclass{article}

\usepackage{etex}
\usepackage{tikz}
\usetikzlibrary{decorations,shapes} 

\usepackage{tikz-cd}
\tikzcdset{
  shift left/.default=+0.7ex,
  shift right/.default=+0.7ex}

\newenvironment{tikzar}[1][]{{}\kern-4pt\begin{tikzcd}[ampersand replacement=\&,#1]}%
{\end{tikzcd}\kern-4pt{}}

\usepackage{enumitem}
\setlist[itemize]{noitemsep, topsep=0pt}

\usepackage{amsmath,stmaryrd}
\usepackage{amsfonts,amssymb}
\usepackage{pstricks}
\usepackage{amsthm}
\usepackage{rotating}

\usepackage[LGR,T1]{fontenc}

\newcommand{\Proc}{\SSS}
\newcommand{\Pred}{\PPP}
\newcommand{\Agnt}{\AAA}
\newcommand{\Term}{\TTT}
\newcommand{\Obsv}{\OOO}
\newcommand{\pr}{{\sf W}}
\newcommand{\kn}{{\sf K}}
\newcommand{\Keys}{\KKK}
\newcommand{\Msgs}{\MMM}
\newcommand{\Cyph}{\CCC}
\newcommand{\Seed}{\RRR}
\newcommand{\Gen}{{\sf Gen}}
\newcommand{\Enc}{{\sf Enc}}
\newcommand{\Dec}{{\sf Dec}}
\newcommand{\DDH}{{\sf DHd}}
\newcommand{\Att}{{\sf Att}}

\newcommand{\tria}[4]{{#1}\ \{{#2}\}_{#3}\ {#4}}

\usepackage{hyperref}

\newcommand{\AAA}{{\cal A}}

\newcommand{\CCC}{{\cal C}}

\newcommand{\JJJ}{{\cal J}}
\newcommand{\KKK}{{\cal K}}

\newcommand{\MMM}{{\cal M}}

\newcommand{\OOO}{{\cal O}}
\newcommand{\PPP}{{\cal P}}

\newcommand{\RRR}{{\cal R}}
\newcommand{\SSS}{{\cal S}}
\newcommand{\TTT}{{\cal T}}

\newcommand{\XXX}{{\cal X}}

\renewcommand{\Bbb}{\mathbb}

\newcommand{\GGg}{{\Bbb G}}

\newcommand{\RRr}{{\Bbb R}}

\newcommand{\ZZz}{{\Bbb Z}}

\mathcode`\<="4268 
\mathcode`\>="5269 
\mathchardef\gt="313E 
\mathchardef\lt="313C 
\newsavebox{\barr}
\savebox{\barr}{\hspace*{-9.5pt}\raisebox{1.25pt}{$
\scriptscriptstyle%
|$}\hspace*{4.5pt}} 
\newsavebox{\barrleft}
\savebox{\barrleft}{\hspace*{-8.5pt}\raisebox{1.25pt}{$
\scriptscriptstyle%
|$}\hspace*{10pt}}

 \def\pushright#1{{
    \parfillskip=0pt            
    \widowpenalty=10000         
    \displaywidowpenalty=10000  
    \finalhyphendemerits=0      
    \leavevmode                 
    \unskip                     
    \nobreak                    
    \hfil                       
    \penalty50                  
    \hskip.2em                  
    \null                       
    \hfill                      
    {#1}                        
    \par}}                      

 \def\qed{\pushright{$\square$}\penalty-700 \smallskip}

\newenvironment{prf}[1]{\begin{trivlist} \item[{\bf ~Proof}#1.]}%
{\qed\end{trivlist}}

\newcommand{\be}[1]{\begin{#1}}

\newcommand{\beq}{\begin{equation}}
\newcommand{\eeq}{\end{equation}}
\newcommand{\ba}[1]{\begin{array}{#1}}
\newcommand{\ea}{\end{array}}
\newcommand{\bea}{\begin{eqnarray}}
\newcommand{\eea}{\end{eqnarray}}
\newcommand{\bear}{\begin{eqnarray*}}
\newcommand{\eear}{\end{eqnarray*}}
\newcommand{\bpr}{\begin{prf}{}}
\newcommand{\epr}{\end{prf}}
\newcommand{\bprf}[1]{\begin{prf}{#1}}
\newcommand{\eprf}{\end{prf}}

\theoremstyle{plain}
\newtheorem{thm}{Theorem}[section]
\newtheorem{prop}[thm]{Proposition}

\newtheorem{defn}[thm]{Definition}
\theoremstyle{remark}

\renewcommand{\to}{\longrightarrow}

\newcommand{\tto}[1]{\xrightarrow{#1}}

\title{Probabilistic annotations for protocol models\\
{\Large \em --- Dedicated to Joshua Guttman --- }
}
\author{Dusko~Pavlovic\thanks{Partially supported by NSF and AFOSR.}\\
\texttt{dusko@hawaii.edu}\\
University of Hawaii, Honolulu HI, USA}

\date{}

\begin{document} 
\maketitle

\begin{abstract}
We describe how a probabilistic Hoare logic with localities can be used for reasoning about security. As a proof-of-concept, we analyze Vernam and El-Gamal cryptosystems, prove the security properties that they do satisfy, and disprove those that they do not. We also consider a version of the Muddy Children puzzle, where children's trust and noise are taken into account.
\end{abstract}

\section{Introduction}
When it was first suggested that I should study security protocols, it was with a remark that the problem was largely solved and that I should simply look for a way to apply the solution to a particular protocol of interest, which happened to be one of the proposals for the IPSec suite. I found a paper that was circulating under the title \emph{'How to solve any protocol problem'} \cite{Widgerson:any}, and spent some time studying the methods of multi-party computation described in it. When I realized that I was not making any progress towards analyzing the IPSec protocol at hand, I went back and found out that the suggested solution of all protocol problems was not the multi-party computation, but strand spaces \cite{GuttmanJ:JAR12,GuttmanJ:JCS14,Guttman:strand-CSFW98,Guttman:strand-CSFW99,Guttman:strand-JCS}. I drew the strand space bundles corresponding to the IPSec proposal the same afternoon. 

Trying to save the science of security protocol design and analysis from its foretold demise, I spent a good part of the next 10 years looking for problems that could not be solved using the strand space model. Each time, I would then meet Joshua Guttman over dinner, usually at one of the Protocol eXchange meetings, and told him that there was this conceptual mismatch between his model and the reality, and he would then suggest how the problem of reality could be adjusted to match the strand spaces, and transformed towards a solution. On one or two occasions when I was too far down the road towards different solutions, I avoided asking about the details. 

But here is a record of something that definitely cannot be done by strand spaces.  It has been clear from the outset that the strand space bundles can be annotated by Floyd-Hoare-style logical annotations \cite{PavlovicD:JCS05,PavlovicD:JCS04,PavlovicD:CSFW01}, and that the various forms of dynamic and epistemic logics, worked out for reasoning about the preconditions, postconditions, and invariants of computations, can be elevated and generalized for reasoning about protocol security \cite{PavlovicD:CSFW05,PavlovicD:ESORICS04} and about the higher-order properties of distributed systems and network interactions \cite{PavlovicD:ICDCIT12}.  But what if we need to reason about the guessing chances, and have to go beyond the Dolev-Yao type of   models \cite{Dolev-Yao}\footnote{There are, of course, many ways to go beyond the Dolev-Yao models and formalize probabilistic and computational reasoning in cryptography. One of the reviewers suggests that Easycrypt \cite{easycrypt-tutorial} should be mentioned. The point here is, however, to try to extend by probabilities the usual Floyd-Hoare annotations, which naturally fit with strand spaces.}?

\section{Crypto-logical systems}
Towards a definition of a {\em crypto-logical system}, we begin from two basic data types: {\em states\/} $\Proc$ and {\em predicates\/} $\Pred$. It is assumed that they are generated by a stratified set of algebraic operations, which allow us to write programs that lead to the states in $\Proc$, and to specify the resulting properties in $\Pred$. In particular, both $\Proc$ and $\Pred$ are built over the same algebra $\Term$ of terms, usually multisorted, assumed to contain enough variables, constants, and function symbols to specify keys, nonces, encryptions, decryptions, hashes, etc. These terms are computed, sent, and received by some actions that may be recorded in $q, s\ldots \in \Proc$, while they may be compared, tested, and reasoned about in $\varphi, \psi\ldots \in \Pred$. 

\subsection{Crude and overly general definition} \label{Sec:Crude}
Given a state space $\Proc$, an algebra of predicates $\Pred$, a set of agents $\Agnt$, and a lattice of observations $\Obsv$, a\/ {\em crypto-logical system}  is defined by the following data:
\begin{itemize}
\item a family of {\em semantic\/} maps
\bear
\Pred \times \Proc \times \Pred & 
\tto{-\{-\}_{A} -}
& \Obsv
\eear
indexed over the agents $A\in \Agnt$, 
\item a measure 
\bear
\Proc &  \tto{\mu} & \RRr_+
\eear 
given with a decomposition  of $\Proc$ into a disjoint union $\Proc = \coprod_{i\in I} \Proc_i$ of unit sets $\Proc_i$, i.e. such that $\Pr(\Proc_i) = 1$ holds  for each $i\in I$. (Each restriction $\Pr_i$ of $\Pr$ to $\Proc_i$ is thus a probability measure.)
\end{itemize}

\paragraph{Remarks.} The above definition is more general than will be needed in this paper. But it conveys the big picture and the general path.  

First of all, we do not need an abstract lattice of observations $\Obsv$, but will always take $\Obsv = \{0,1\}$, and work with the usual Hoare triples $\varphi \{q\}_A \psi$, which are simply the elements of a ternary relation over $\Pred \times \Proc \times \Pred$. The reason for the above formulation is that the probabilistic analysis below will suggest that the {\em probabilistic\/} Hoare triples, evaluated in $\Obsv = [0,1]$, are also of interest, and in fact simplify some aspects of the reasoning. This option should be kept in mind for future work.

The decomposition of the state space $\Proc = \coprod_{i\in I} \Proc_i$ allowing the decomposition of the measure $\mu$ into the probability measures $\mu_i$ will also not play a significant role. It is in principle needed in the examples in Sections \ref{Sec:def} and \ref{Sec:example}, where the state spaces will be certain powers of the monoid $\Sigma = \{0,1\}^\ast$ of bitstrings, decomposed into $\Sigma = \coprod_{n=0}^\infty \{0,1\}^n$, with the uniform probability distribution over each finite component $\{0,1\}^n$. But this is spelled out in many textbooks, and the decomposition would flood the notations by information that is inessential for this paper, and hide the aspects that are essential. So we reduce the measure $\mu: \Proc   \to \RRr_+$ to the component probabilities $\Pr:\Proc\to [0,1]$, omitting the indices as they are easily reconstructed  in all cases.

Furthermore, a state $q$ in the space $\Proc$ may or may not contain a record of a particular computation, run, or process that led to it. Short of a better word, we stretch the word "state" to mean "a result of a computation" --- whatever part of it we may choose to record.  Sometimes it may be the whole history, even including the intermediary results; sometimes just the outcome. A consequence is that $\Proc$ may be closed under the usual programming and process operations, or it may be structured by the recorded data alone. In the former case, the usual rules of the Hoare logic will apply. In the latter case, when the concrete computations are not reflected by  modal operators in $\Proc$,
the Hoare notation boils down to
\bear
q\underset{A}{\models} \psi & \iff & \tria \top q A \psi
\eear
On the other hand, when the preconditions do play an essential role, relying upon the Hoare logic tradition and intuition seems appropriate, and useful. 

In any case, we always require that the semantic maps $\Pred\times \Proc \times \Pred \to \Obsv$ preserve the lattice structure of $\Pred$, contravariantly in the first,  precondition argument, and covariantly in the postcondition.

\subsection{Information sets and preorders of states} 
We say that, for an agent $A$, a process $q'$ {\em refines\/} a process $q$, or that it {\em contains more information than $q$}, and we write $q\underset{A}{\sqsubseteq} q'$, whenever $q'$ satisfies, as far as $A$ can tell, all the requirements that $q$ satisfies:
\bear
q\underset{A}{\sqsubseteq} q' & \iff& \forall \varphi\psi.\ \varphi\{q\}_A \psi \leq \varphi\{q'\}_A \psi
\eear
Two processes are indistinguishable for the agent $A$ if they satisfy the same requirements
\bear
q\underset{A}{\sim} q' & \iff& q\underset{A}{\sqsubseteq} q'\ \wedge\ q\underset{A}{\sqsupseteq} q'\\
&\iff & \forall \varphi\psi.\ \varphi\{q\}_A \psi = \varphi\{q'\}_A \psi
\eear
The $\underset{A}{\sim}$-equivalence classes are $A$'s {\em information sets}. The quotient $\Proc_A = \Proc/\underset{A}{\sim}$ is $A$'s information view. $A$'s information set at $q$ is written $q_A\in \Proc_A$.

\subsection{Refining the definition of crypto-logical systems} 
The data type $\Pred$ of predicates is assumed to support the usual logical connectives, which make it into a lattice. Moreover, it is also closed under a family of modalities $\pr_\iota$, indexed over some \emph{subjective evaluations}\/ $\iota\in  \JJJ[0,1]$, which will be just numbers between 0 and 1 in the simple examples below, but need to be generalized for some more involved cryptographic constructions. Semantics of these logical operations is defined by the following conditions
\bea
\tria{(\varphi_1\vee \varphi_2)} {q} {A} {\psi} & \iff & (\varphi_1\{q\}_A \psi)\ \wedge\ (\varphi_2\{q\}_A \psi) \label{disj}\\
\tria \varphi q A {(\psi_1\wedge \psi_2)} & \iff & (\varphi\{q\}_A \psi_1)\ \wedge\ (\varphi\{q\}_A \psi_2) \label{conj}\\
\tria \varphi q A {(\pr_\iota \psi)} & \iff & \Pr\left(\tria \varphi {s} A \psi\ \Big|\ s\underset{A}{\sim} q\right) \in \iota \label{pr}
\eea
The Hoare triples here are the standard ones, evaluated in $\Obsv = \{0,1\}$, as explained in  Sec.~\ref{Sec:Crude}. Clause \eqref{pr} extends the standard Hoare logic for probabilistic reasoning. The idea is that  
\begin{itemize}
\item $A$'s subjective probability that $\psi$ holds after $\varphi$ at $q$ is equal to
\item the objective probability that $\psi$ holds after $\varphi$ at a randomly chosen state $s\underset{A}\sim q$.
\end{itemize}
By definition, the conditional probability in the last clause unfolds to
\bear
\Pr \left(\tria \varphi {s} A \psi \ \ \Big| \ \ s\underset{A}{\sim} q\right) & = &  \frac{\Pr\left\{s\in \Proc\ |\ s\underset{A}{\sim} q\ \wedge\ \tria \varphi {s} A \psi \right\}}{\Pr\left\{s\in \Proc\ |\ s\underset{A}{\sim} q\right\}} 
\eear

\paragraph{The subjective vs objective probability conundrum} goes back to the earliest days of probability theory \cite{prob-subj-obj}, and persists as a useful distinction even in cryptographic reasoning. The objective probability is a number, which can be obtained, e.g., by counting frequencies. An observer of a random process, however, may only be able to estimate that a probability falls within a certain interval, or just in a set, measurable modulo computational indistinguishability. There are thus various generality levels at which the family  $\JJJ[0,1]$ of subjective evaluations may need to be modeled. To capture the standard cryptographic definitions in Sec.~\ref{Sec:def}, the subjective evaluations from $\JJJ[0,1]$ will need to be feasibly computable subintervals of $[0,1]$. For the simple examples presented in Sec.~\label{Sec:example}, on the other hand, rational numbers will suffice.

\subsection{Probability vs. knowledge}
Note that the statement $\pr_1\psi$, saying that $\psi$ is satisfied with probability 1,
\bear
\tria \varphi q A {(\pr_1 \psi)} & \iff & \Pr\left(\tria \varphi {s} A \psi\ \Big|\ s\underset{A}{\sim} q\right) = 1
\eear
can be viewed as a generalization of the knowledge modality $\kn \psi$ for $A$ defined by
\bear
\varphi\ \{q\}_A  {\left(\kn \psi\right)} & \iff & \forall s\in \Proc.\  s\underset{A}{\sim} q\Rightarrow \varphi\{s\}\psi
\eear
where the logical implication $s\underset{A}{\sim} q\Rightarrow \varphi\{s\}\psi$ is replaced by the stochastic implication 
\bear
\big[ s\underset{A}{\sim} q\Rightarrow \varphi\{s\}\psi \big] & \iff &  \Pr\left(\tria \varphi {s} A \psi\ \Big|\ s\underset{A}{\sim} q\right) = 1
\eear
Intuitively, this stochastic implication says that the implication is valid almost everywhere, i.e. everywhere except at a set of measure 0. While the usual semantics of knowledge tells that $\kn \psi$ is satisfied  for $A$ after $\varphi$  at $q$ if $\psi$ is satisfied after $\varphi$ at every $s\underset{A}{\sim} q$, the probabilistic knowledge $\pr_1 \psi$ is satisfied after $\varphi$ for almost all $s\underset{A}{\sim} q$, i.e. with a possible exception of a set of measure 0. For each $A$, the statements $\kn \psi$ and $\pr_1 \psi$ are {\em almost everywhere\/} equivalent, i.e. they only differ at a set of states of measure 0. Since cryptographic proofs are not just up to sets of measure 0, but usually identify even the ensembles that are computationally indistinguishable\footnote{Two ensembles are computationally indistinguishable when their differences cannot be detected by polynomially bounded computations, e.g. because they occur only superpolynomially far down the strings of digits of their probabilities.} the knowledge modality should, for all cryptographic purposes, be identified with $\pr_1$.

\subsection{Global semantics} 
We say that a requirement is satisfied globally if some agent observes that it is satisfied
\bea\label{glob}
\tria \varphi q {} \psi & \iff & \exists X.\ \tria \varphi q X \psi
\eea
In practice, crypto-logical systems are often given by 
\begin{itemize}
\item a {\em global semantics}
\bear
\Pred \times \Proc \times \Pred & 
\tto{-\{-\} -}
& \Obsv
\eear
\item a family of {\em views\/}
\bear
\Proc & \tto{(-)_A} & \Proc
\eear 
indexed by $A\in \Agnt$ such that 
\bea 
(\forall X.\  q_X = q'_X) & \iff &  q = q'\mbox{ and} \label{one}\\
 \tria \varphi q {} \psi & \iff & \exists X.\ \tria \varphi {q_X} {} \psi \label{two}
 \eea
\end{itemize}
Local semantics can then be defined by
\bear
\tria \varphi q A \psi &\iff & \tria \varphi {q_A} {} \psi
\eear
Condition (\ref{two}) implies that (\ref{glob}) recovers the global semantics. Condition (\ref{one}) implies that $q\sim_A q' \iff q_A = q'_A$. In other words, since all $q' \sim_A q$ satisfy the same requirements $\varphi \{q'\}_A \psi$ if and only if $q_A$ satisfies them, then $q_A$ can be taken as the canonical representative of the information set $[q]_A\in \Proc_A$.

\subsection{Knowledge of probability vs probability of knowledge}
The {\em logical\/} interpretation of the probabilistic modality $\pr_\iota$, proposed in (\ref{pr}), was stated over the observations in $\Obsv = \{0,1\}$. Allowing the observations to be evaluated in $\Obsv = [0,1]$, and replacing the logical equivalence in (\ref{disj}) and (\ref{conj}) by the equality or indistinguishability of probabilities, leads to the probabilistic interpretation of the knowledge modality
\bear
\varphi\ \{q\}_A  {(\kn \psi)} & = & \Pr\left(\varphi\ \{s\}\ \psi\ \Big| \ s\underset{A}{\sim} q\right) 
\eear
and promotes $\pr_\iota$ into a \emph{confidence}\/ modality
\bear
\varphi\ \{q\}_A  {(\pr_\iota \psi)} & = & \Pr\bigg(\Pr\left(\varphi\ \{s\}\ \psi\ \Big| \ s\underset{A}{\sim} q\right)\in \iota\bigg) 
\eear
But this refined view has to be left for future work, as it requires first spelling out the standard view of familiar concepts, which barely fit in the rest of this paper.



\section{Cryptographic definitions in crypto-logic}\label{Sec:def}
A {\em cryptosystem\/} consists of three agents, each executing a single probabilistic algorithm:
\begin{itemize}
\item key generation $\Gen : \Seed \to \Keys\times \Keys$, 
\item encryption $\Enc : \Keys \times \Seed \times \Msgs \to \Cyph$, and
\item decryption $\Dec : \Keys\times \Cyph\to \Msgs$,
\end{itemize}
such that
\bear
\Dec(\overline{k},\Enc(k,x,m)) & = & m
\eear
where $<k,\overline{k}> = \Gen(y)$ for some $y\in \Seed$.
Here $\Seed$ represents the data type of random seeds, $\Keys$ is the datatype of keys, $\Msgs$ the datatype of plaintext messages, and $\Cyph$ the ciphertexts. All datatypes are assumed to be finite, although unfeasibly large, so that it is sometimes convenient to assume that they are countably infinite. Each of them is given with a frequency measure
\bear
\Pr & :& \XXX \to [0,1]
\eear
When no confusion seems likely, we shall denote a random variable sampling from $\XXX$ also by $\XXX$, and write $\Pr(x\in \XXX)$ where most probability theory textbooks would write $\Pr(\XXX = x)$.

Besides the principals of the cryptosystem, a definition of a security property that it may satisfy involves an attacker $\Att$, which may operate any number of algorithms.

\paragraph{Remark.} The notion of an {\em algorithm}\/ is used here in the broadest sense, accomodating the various notions of computation.  While the {\em computational\/} notions of security are defined assuming Probabilistic Polynomial-time Turing (PPT) machine as the standard model of computation, the {\em information-theoretic\/} security is defined over a notion of computation which boils down to mere {\em guessing}\/ (of a message, a key, etc.), according to given frequency distributions. We begin with an information-theoretic definition.

\begin{defn} A cryptosystem is\/ {\em perfectly secure} if Attacker's chance to guess a message $m$ at a state $C$, when he is given a ciphertext $c = E(k,x,m)$ is the same as his chance to guess that message at a state O, where he is not given any data, and can just randomly sample the space $\Msgs$ of messages:
\be{align}
C \models \pr_\iota (m\in \Msgs) && \iff  && O \models \pr_\iota (m\in \Msgs) \tag{IT-SEC}
\end{align}
\end{defn}

\begin{defn} {\em Semantic\/} (or {\em chosen plaintext\/}) security of a cryptosystem is tested by the following protocol:
\begin{itemize}
\item the Attacker computes (or randomly selects) two messages, $m_0$ and $m_1$, and sends them to the Encryption oracle;
\item the Encryption oracle tosses a coin, i.e. randomly selects a bit $b$, and a seed $x\in\Seed$, computes the ciphertext $c = E(k,x,m_b)$, and sends it to the Attacker.
\end{itemize}
The cryptosystem is semantically secure if Attacker's chance to compute (or to guess) the bit $b$ at the final state $C$, when $c$ is known to him, is not greater than his chance to guess $b$ at the initial state $O$, without any data, i.e.
\be{align}
C \models \pr_\iota \left(b=1\right) && \iff &&  O \models \pr_\iota \left(b =1\right) \tag{IND-CPA}
\end{align}
\end{defn}

\begin{defn} {\em Adaptive\/} (or {\em chosen ciphertext\/}) security of a cryptosystem is tested by the following protocol:
\begin{itemize}
\item the Attacker computes (or randomly selects) two messages, $m_0$ and $m_1$, and sends them to the Encryption oracle;
\item the Encryption oracle tosses a coin, i.e. randomly selects a bit $b$, and a seed $x\in\Seed$, computes the ciphertext $c = E(k,x,m_b)$, and sends it to the Attacker,
\item the Attacker is then allowed to consult the Decryption oracle, to obtain the decryption $d = D(\overline{k},c')$, of a chosen piece if ciphertext $c'$ is feasibly constructed from $m_0, m_1$ and $c$, but differs from $c$, i.e. $c'\neq c$.
\end{itemize}
The cryptosystem is adaptive secure if Attacker's chance to compute (or to guess) the bit $b$ at the final state $C$, when the ciphertext $c$ and the decryption $d$ are known to him, is not greater than his chance to guess $b$ at the initial state $O$, without any data, i.e.
\be{align}
C \models \pr_\iota \left(b=1\right) && \iff  &&O \models \pr_\iota \left(b =1\right) \tag{IND-CCA}
\end{align}
\end{defn}

\paragraph{Remark.} Varying the notion of computation in the above definition results in different notions of security. If the notion of computation is reduced to guessing, i.e. if the Attacker can only randomly choose $m_0$ and $m_1$, and only randomly guess $b$, but possibly following a probability distribution skewed by the knowledge of $c$, then we get a weaker notion of security than the one where the Attacker can perform more structured computation, e.g. of a Probabilistic Polynomial-Time Turing Machine (PPT).

\section{Examples of reasoning in crypto-logic}\label{Sec:example}
\subsection{Security of the Vernam cryptosystem}
In the Vernam cryptosystem, we take
\bear
\Keys & = & \{0,1\}^\ell\\
\Msgs & = & \Keys^j\\
\Cyph & = & \Msgs\\
\Seed & = & 1
\eear
and then define
\bear
E(k,m)\ = \ D(k,m) & = & k^j\oplus m
\eear
where $\oplus$ is the {\em exclusive or} operation, and $k^j$ is the $j$-tuple concatenation of a key $k$. We assume that the messages have a fixed number of blocks $j$ just to avoid inessential notational details. The probability distributions over $\Keys$ and over $\Msgs$ are given, and they determine
\bear
\Pr (c\in \Cyph) & = & \sum_{k^j \oplus m = c} \Pr(m\in \Msgs) \cdot \Pr(k\in \Keys) 
\eear

The Vernam cryptosystem is called {\em one-time pad\/} when $j=1$, i.e. when a key is used to encrypt just one block.

\begin{prop} One-time pad is perfectly secure. The Vernam cryptosystem is not perfectly secure for $j\geq 2$.
\end{prop}

\bpr
To model the (IT-SEC) testing of the Vernam cryptosystem, we use  as the states in $\Proc$ the substrings of the triples $<k,m,c> \in \Keys \times \Msgs \times \Cyph$, subject to the constraint that $c = k^j \oplus m$. Each state can be construed as the record of an encryption session, where the key $k$ is first generated and sent from $\Gen$ to $\Enc$, then the message $m$ is chosen and encrypted by $\Enc$ into $c=k\oplus m$, and finally, the ciphertext $c$ is sent to $\Dec$ and $\Att$.

For each agent $X\in \{\Gen,\Enc,\Dec,\Att\}$ we define the view function $\Proc \tto{(-)_X} \Proc$ to be
\bear
<k,m,c>_\Gen & = & <k>\\
<k,m,c>_\Enc & = & <k,m,c>\\
<k,m,c>_\Dec & = & <k,c>\\
<k,m,c>_\Att & = & <c>
\eear

The data type $\Pred$ of predicates is generated from the formulas of binary arithmetic, extended with the probabilistic modalities $\pr_\iota$. 

We define semantics by stipulating that $\tria \varphi q {X} \psi$ is satisfied  whenever the implication $\varphi(q_X)\Rightarrow \psi(q_X)$ is provable in binary arithmetic and elementary probability theory, starting from the given distributions $\Pr_\Msgs$, and $\Pr_\Keys$.

Towards a proof of (IT-SEC) property for $j=1$, first note that
\bear
<> \models \pr_a (m\in \Msgs) &\quad \iff \quad & \Pr(m\in \Msgs) = a \\
<c> \models \pr_b (m\in \Msgs) & \iff & \Pr(m\in \Msgs\ |\ c\in \Cyph) = b
 \eear
On the other hand,
\bear
\Pr(m\in \Msgs\ |\ c\in \Cyph) & = & \frac{\Pr(c\in \Cyph\ |\ m\in \Msgs)\cdot \Pr(m\in\Msgs)}{\Pr(c\in\Cyph)} \\
& = & \Pr(m\in \Msgs)
\eear
holds because
\bear
\Pr(c \in \Cyph\ |\ m \in \Msgs) & = & \Pr(c=k\oplus m \in \Cyph\ |\ m \in \Msgs)\\
& = &  \Pr(k = c\oplus m \in \Keys\ |\ m \in \Msgs) \\
& = &  \Pr(k \in \Keys)
\eear
and
\bear
\Pr(c \in \Cyph) & = & \sum_{m\in \Msgs} \Pr(c \in \Cyph\ |\ m \in \Msgs) \cdot \Pr(m\in \Msgs)\\
& = &  \Pr(k \in \Keys) \sum_{m\in \Msgs} \Pr(m\in \Msgs) \\
& = & \Pr(k \in \Keys)
\eear
It follows that $<> \models \pr_a (m\in \Msgs)$ and  $<c> \models \pr_b (m\in \Msgs)$ are satisfied if and only if $a = b$.

For the Vernam cipher with $j\geq 2$, the probability $\Pr(c\in \Cyph\ |\ m\in \Msgs)$ does not boil down to $\Pr(k\in \Keys)$. Given $m = m_1 :: m_2 :: \cdot m_j$, then $c$ must be in the form $c = c_1 :: c_2 :: \cdot c_j$ where $c_1\oplus m_1 = c_2 \oplus m_2 =\cdots =c_j\oplus m_j$ equals the key $k$. For $c\in \Cyph$ which are not in that form, $\Pr(c\in \Cyph\ |\ m\in \Msgs) = 0$. For those that are, $\Pr(c\in \Cyph\ |\ m\in \Msgs) = \Pr(k\in \Keys)$ remains valid. By a similar reasoning, 
\bear
\Pr(m\in \Msgs\ |\ c\in \Cyph) & = & \begin{cases} \Pr(k\in \Keys) & \mbox{ for } m_1\oplus c_1 = \cdots = m_j\oplus c_j \\
0 & \mbox{ otherwise}
\end{cases}
\eear
This shows that the Vernam cryptosystem does not satisfy (IT-SEC) for $j\geq 2$.
\epr

\begin{prop}\label{Prop:ver-cpa}
If a Vernam cryptosystem is used to encrypt even one bit more than one block, then it is not semantically (IND-CPA) secure, i.e. it can be broken by a chosen-plaintext attack. 
\end{prop}

\paragraph{Remark.} Note that Attacker's capability to choose a plaintext is computational, and not just stochastic: they can determine the structure of the messages $m_0$ and $m_1$ in the CPA-test, and not just rather than just randomly sample from some source.

\bprf{ of Prop.~\ref{Prop:ver-cpa}}
To model the Vernam cryptosystem where one bit more than one block is encrypted, we take
\bear
\Msgs\ = \ \Cyph\  & = &\  \Keys\times \{0,1\}
\eear
To model the (IND-CPA) testing of this cryptosystem, we use as the states in $\Proc$ the substrings of the triples $<k,m_0,m_1,b,c> \in \Keys \times \Msgs^2 \times \{0,1\} \times \Cyph$, subject to the constraint that $c = k' \oplus m$, where $k' = k :: k_0$ is the key $k$ with the first bit repeated at the end. Each state can be construed as the record of an encryption session, where the key $k$ is first generated by $\Gen$, and securely conveyed to $\Enc$ and $\Dec$, while on the other side the messages $m_0,m_1$ are generated by $\Att$ and sent to $\Enc$, who then chooses the bit $b$, computes the ciphertext $c =k^j\oplus m_b$ and sends it $c$ to $\Dec$ and $\Att$.

For each agent $X\in \{\Gen,\Enc,\Dec,\Att\}$ we define the view function $\Proc \tto{(-)_X} \Proc$ to be
\bear
<k,m_0,m_1,b,c>_\Gen & = & <k>\\
<k,m_0,m_1,b,c>_\Enc & = & <k,m_0,m_1,b,c>\\
<k,m_0,m_1,b,c>_\Dec & = & <k,c>\\
<k,m_0,m_1,b,c>_\Att & = & <m_0,m_1,c>
\eear
The data type of predicates $\Pred$ and the semantics of $\tria \varphi q {X} \psi$ are just like in the proof of the preceding proposition.

Towards a proof that (IND-CPA) not satisfied, we note that
$<> \models \pr_{\frac{1}{2}} \left(b = 1\right)$ holds, because\footnote{We assume that the coin is fair. If it is biased, the argument goes through for any probability $p$ instead of $\frac{1}{2}$, provided that  $p \neq 0$ and $p\neq1$.}  $\Pr(b=1) = \frac{1}{2}$.

On the other hand, we show that the attacker can construct the messages $m_0$ and $m_1$ in such a way that
$<c> \models \pr_1 (b=1)$ holds if and only if $c = k' \oplus m_1$, and otherwise  $<c> \models \pr_0 (b=1)$ holds. Either way, $<c> \models \pr_{\frac{1}{2}} (b=1)$  does not hold, which implies that
\bear
\left(<c> \models \pr_{\frac{1}{2}} (b=1)\right) &\quad\ \   \not\!\!\!\Longleftrightarrow \quad&  \left(<> \models \pr_{\frac{1}{2}} (b =1)\right)
\eear

Towards the counterexample for (IND-CPA), let
\bear
m_0 & = & 0^\ell :: 0\\
m_1 & = & 0^\ell :: 1
\eear
which gives 
\bear
c_0 & = & k_0 :: k_1:: \cdots :: k_0\\
c_1 & = & k_0 :: k_1:: \cdots :: \neg k_0
\eear
and
\bear
<c_0> & \models & \pr_0(b = 1)\\
<c_1> & \models & \pr_1(b=1)
\eear
\epr

\subsection{El-Gamal}
Let $\GGg$ be a cyclic group\footnote{Here we hide away some details. $\GGg$ is usually taken to be a cyclic subgroup of the multiplicative group of a field $\ZZz_p$. But while the reader familiar with the system, or a student of any cryptography textbook, will have no trouble recovering the details swept under the carpet, carrying them around here would distract from the main idea.} of order $n$ with a generator $g$. In other words, the elements of $\GGg$ can be listed in the form $g, g^2, g^3, \ldots, g^{n-1}, 1$. The types of the El-Gamal cryptosystem are taken to be
\bear
\Keys &  = & \GGg \times \ZZz_n \\
\Seed & = & \ZZz_n\\
\Msgs & = & \GGg \\
\Cyph & = & \GGg \times \GGg
\eear
The keys $<k,\overline{k}> = \Gen(a)$ are set to be
\bear
k & = & g^a\\
\overline{k} & = & a
\eear
and the encryption and decryption functions are
\bear
E(k,r,m) & = & <g^r, k^r\cdot m>\\
D(\overline{k},c) & = & \frac{c_2}{c_1^{\overline{k}}}
\eear
where $c = <c_1,c_2>$. This defines a cryptosystem because
\[
D\left(\overline{k},E(k,r,m)\right)\ =\  \frac{k^r \cdot m}{\left(g^{r}\right)^{\overline{k}}} \ 
= \ \frac{g^{a\cdot r} \cdot m}{g^{r\cdot a}} \ = \  m
\]

\begin{defn}\label{Def:DDH}
The \emph{Diffie-Hellman decision} is the predicate $\DDH : \GGg^3 \to \{0,1\}$ defined by 
\bear
\DDH(x,y,z) & \iff &  \exists a, b \in \ZZz_n.\ x = g^a\wedge y = g^b \wedge z = g^{ab}
\eear 
where we abbreviate $\DDH(x,y,z) = 1$ to $\DDH(x,y,z)$, and write $\neg \DDH(x,y,z)$ when $\DDH(x,y,z) = 0$. The \emph{Decision Diffie-Hellman problem}\/ concerns the guessing algorithms for the Diffie-Hellman decision, i.e. the feasible algorithms with random seeds. The problem is that an algorithm should do better than a coin flip, and output more than half true decisions for a given length of the seeds. Formally, this means that for all $a, b\in \ZZz_n$ a $\DDH$ algorithm should satisfy\footnote{It is required that the chance of $\DDH\left(g^a, g^b, g^{ab}\right) = 1$ is feasibly distinguishable from $\frac 1 2$, i.e. greater by a feasible function. It follows that the chance of $\DDH\left(g^a, g^b, g^{d}\right) = 1$ for $d\neq ab$ is also significantly smaller than $\frac 1 2$ by a feasible function.}
\bear
\Pr\Big(\DDH\left(g^a, g^b, g^{ab}\right)\Big) & \gt & \frac 1 2 
\eear 
The \emph{Decision Diffie-Hellman (DDH) assumption}\/ is that the Diffie-Hellman problem has no solution, i.e. that no feasible algorithm for guessing the Diffie-Hellman decision can do better than the coin flip.
\end{defn}

\begin{prop}
The El-Gamal cryptosystem is semantically secure if and only if the Decision Diffie-Hellman assumption is true.
\end{prop}

\bpr
To model the (IND-CPA)-testing of the El-Gamal cryptosystem, i.e. choosing the plaintexts that will yield distinguishable ciphertexts, we use  as the states in $\Proc$ the substrings of the tuples 
\[ \left<<k,\overline{k}>,r,m_0,m_1,b,c\right> \in \Keys \times \Seed  \times \Msgs^2   \times\{0,1\}\times \Cyph\]
where $k = g^{\overline{k}}$ and $c = \left< g^r, k^r\cdot m_b\right>$. Each state can be construed as the record of a testing session, where the keys $\overline{k}$ and $k$ are generated, the first one is sent from $\Gen$ to $\Enc$, the second one is announced publicly; the messages $m_0, m_1$ are chosen and sent from $\Att$ to $\Enc$, the bit $b$ and the ciphertext $c$ are generated and sent from $\Enc$ to $\Att$ and $\Dec$.

For each agent $X\in \{\Gen,\Enc,\Dec,\Att\}$ we define the view function $\Proc \tto{(-)_X} \Proc$ to be
\bear
\left<k,\overline{k},r,m_0,m_1,b,c\right>_\Gen & = & \left<k,\overline{k}\right>\\
\left<k,\overline{k},r,m_0,m_1,b,c\right>_\Enc & = & \left<k,r,m_b,c\right>\\
\left<k,\overline{k},r,m_0,m_1,b,c\right>_\Dec & = & \left<\overline{k},m_b,c\right>\\
\left<k,\overline{k},r,m_0,m_1,b,c\right>_\Att & = & \left<k,m_0,m_1,c\right>
\eear

Suppose that for the El-Gamal El-Gamal cryptosystem holds
\be{align}\label{eq:nindcpa}
 C \models \pr_\iota \left(b=1\right) && \not\hspace{-.8em}\iff && O \models \pr_\iota \left(b =1\right) \tag{$\neg$ IND-CPA}
\end{align}
Since for a fair coin (i.e. uniformly distributed) $b\in \{0,1\}$ it is certainly true that $O \models \pr_{\frac 1 2} \left(b =1\right)$. The assumption \eqref{eq:nindcpa} thus means that there is an attack that makes $C \models \pr_{\frac 1 2} \left(b=1\right)$ false.  There are thus algorithms 
\[\Att_0:\Msgs^2\qquad \mbox{ and }\qquad \Att_1: \GGg \times \Msgs^2 \times \Cyph \to \{0,1\}\] such that for $\Att_0 = <m_0,m_1>$ and any $b\in \{0,1\}$ holds
\bear
\Pr\Big(\Att_1\left(k,m_0, m_1, <g^r, k^r\cdot m_b> \right) = b\Big) &\gt & \frac 1 2
\eear
The Diffie-Hellman decision $\DDH(x,y,z)$ can now be computed for any given $x, y$ and $z$ from $\GGg$ as follows:
\begin{itemize}
\item Set and announce the public key to be $k = x$. 
\item Let $\Att_0$ generate and send the messages $m_0$, and $m_1$.
\item Pick any $b\in\{0,1\}$ and announce $c = \left<y, z\cdot m_b\right>$.
\item Set $\DDH(x,y,z) = 1$ if and only if $\Att_1$ correctly guesses $b$.
\end{itemize}
In summary,
\begin{align}
\DDH(x,y,z) &  =  \begin{cases}
1 & \mbox{ if }  \ \ \ \Att_1\left(k,m_0, m_1, <y,z\cdot m_0>\right) = 0 \\
& \mbox{ and } \Att_1\left(k,m_0, m_1, <y,z\cdot m_1>\right) = 1\\
0 & \mbox{ otherwise}
\end{cases} \tag{$\neg$ DDH}
\end{align}
The other way around, assuming ($\neg$ DDH) with a Diffie-Hellman decision algorithm $\DDH$ significantly better than a coin flip, the attacker $\Att_0$ may generate $m_0$ and $m_1$ randomly, since $\Att_1$ can always use $\DDH$ to decide which of the messages has been encrypted
\bear
\Att_1\left(k,m_0, m_1, <c_0,c_1> \right) &  = & \begin{cases}
b & \mbox{ if } \DDH\left(k,c_0, \frac{c_1}{m_b}\right)\\
\bot & \mbox{ otherwise}
\end{cases}
\eear
Checking that this yields ($\neg$ IND-CPA) is straightforward.
\epr

\begin{prop}
The El-Gamal cryptosystem is not adaptively secure, i.e. it can be broken by a chosen ciphertext attack.
\end{prop}

\bpr
To model the (IND-CCA) (chosen ciphertext) testing of the El-Gamal cryptosystem, we use  as the states in $\Proc$ the substrings of the tuples 
\[ \left<<k,\overline{k}>,r,m_0,m_1,b,q,c,c',d\right> \in \Keys \times \Seed  \times \Msgs^2   \times\{0,1\}\times  \Seed\times  \Cyph^2\times \Msgs\]
where $k = g^{\overline{k}}$ and $c= \left<g^r, k^r\cdot m_b\right>$, $c'\neq c$, and $d = D(\overline{k},c')$. The projections can be
\bear
\left<k,\overline{k},r,m_0,m_1,b,q,c,c',d\right>_\Gen & = & \left<k,\overline{k}\right>\\
\left<k,\overline{k},r,m_0,m_1,b,q,c,c',d\right>_\Enc & = & \left<k,r,m_b,c\right>\\
\left<k,\overline{k},r,m_0,m_1,b,q,c,c',d\right>_\Dec & = & \left<\overline{k},c',d\right>\\
\left<k,\overline{k},r,m_0,m_1,b,q,c,c',d\right>_\Att & = & \left<k,m_0,m_1,q,c,c',d\right>
\eear

In order to gain advantage in determining $b$, the Attacker just needs to generate $q\neq 1$, and for $c=<c_1,c_2>$ set $c'= \left<c_1, q\cdot c_2\right>$ . Then $d = q\cdot m_b$, and $b$ can be determined with certainty, by comparing $m_b = \frac{d}{q}$ with $m_0$ and $m_1$.
\epr

\subsection{Towards protocols for noisy muddy mistrustful children}
In some cryptanalytic attacks, the Attacker is a distributed system, consisting of several processes which locally make different observations, and send messages to each other. The Muddy Children Puzzle can be viewed as a rudimentary example of such a situation. An unknown bitstring $m\in \Msgs = \{0,1\}^\ell$ can be thought of as denoting which members of a group of $\ell$ children have a muddy forehead. The fact that each child only sees other children's foreheads, but not its own, corresponds to the fact that an Attacker may consist of $\ell$ observers $\Att_i$, $i=1,\ldots,\ell$, and each $\Att_i$ sees the bits $m_k$ for $k\neq i$ but does not see $m_i$.

In the usual version of the puzzle, the father tells the children that at least one of them has a muddy forehead, and asks each child whether it knows if its forehead is dirty. He asks them in rounds: after they all say "No", he asks them all again, and so on. Using their view of other childrens' foreheads, and hearing their answers, each child can at some point tell whether its forehead is dirty. It is assumed that each child is a perfect reasoner: it will prove everything that can be proved at that point in time. At each point in time, each child either knows with certainty whether his forehead is muddy or does not know it at all. 

In the probabilistic version, each child is trying to estimate the probability that his forehead is muddy. Initially, having finished playing together, the children have an estimate of the distribution $p:n\to [0,1]$, where $p_k$ is the probability that exactly $k$ of them have a dirty forehead. If a child sees $k$ dirty foreheads, then it knows for sure that there are either $k$ or $k+1$ dirty foreheads alltogether. So the initial probability that its own forehead is dirty is $\frac{p_{k+1}}{p_k + p_{k+1}}$.

Like in the usual version, each child then proceeds to announce, in rounds, whether it knows the state of its forehead. Knowing each other, they all also have an estimate of the probability that the statement that each of them is making is false (for one reason or another).

In other words, the Attackers initially know the probability $p_k$ that there are exactly $k$ 1s in $m$. Then each $\Att_i$ is allowed to broadcast to all $\Att$s a message, telling whether he knows $m_i$ or not. These broadcasts continue in rounds. After a finite number of such broadcasts, all $\Att$s can compute all of the bitstring $m$. 

The reasoning that allows this is one of the motivating examples behind knowledge logics. Generalizing the knowledge modality into the probability modality allows refined reasoning, where unreliability of the Attacker's communications can be taken into account: their broadcast bits can be flipped, with a given probability. This probability can be thought of as a measure of noise, or of mistrust among the children. 

\subsection*{To be continued}
While gathering the references, in particular those that I missed during the years of missed Protocol eXchanges, I encountered reports about the extensions of strand spaces, bundles, and shapes that support quantitative and hybrid forms or reasoning about security \cite{GuttmanJ:shapes,GuttmanJ:hybrid,Guttman:metric}. The tradition of Joshua explaining to me how what I presented could be done using the strand space model is hoped to be continued in the future.

\bibliography{PavlovicD,protocol,crypto,prob}
\bibliographystyle{plain}

\end{document}